# Relationship between polytropic index and temperature anisotropy in space plasmas

G. Livadiotis, G. Nicolaou

Southwest Research Institute, San Antonio, TX, USA; glivadiotis@swri.edu

## Abstract

The paper develops a theoretical relationship between the polytropic index and the temperature anisotropy that may characterize space plasmas. The derivation is based on the correlation among the kinetic energies of particles with velocities described by anisotropic kappa distributions. The correlation coefficient depends on the effective dimensionality of the velocity distribution function, having its shape determined by the temperature anisotropies caused by the ambient magnetic field; on the other hand, the effective dimensionality is directly dependent on the polytropic index. This analysis leads to the connection between the correlation coefficient, effective dimensionality of velocity space, and the polytropic index, with the ratio of temperature anisotropy. Moreover, a data and statistical analysis is performed to test the application of the theoretically developed model, in the solar wind proton plasma near 1 AU. The derived theoretical relationship is in good agreement with observations, showing that the lowest and classical value of the adiabatic polytropic index occurs in the isotropic case, while higher indices characterize anisotropic plasmas. Finally, possible extensions of the theory considering (i) non-adiabatic polytropic behavior, and (ii) more general distributions, are further discussed.

## 1. Introduction

The polytropic behavior is a thermodynamic property of the particle system – the plasma. The value of the polytropic index indicates a certain thermodynamic process of the plasma particles, that is, the transition of the plasma from one thermodynamic state to another; the polytropic index determines the heat flux during this transition (e.g., Parker 1963, Chandrasekhar 1967, Verma et al. 1995, Vasquez et al. 2007, Sorriso-Valvo et al. 2007, Marino et al. 2008, Livadiotis 2019a).

The thermal pressure $p$, density $n$, and temperature $T$ of particle systems, driven by a dynamical potential energy with a positional dependence, are characterized by a polytropic relationship, such as

$$p(\vec{r}) \propto n(\vec{r})^\gamma \text{ , or } n(\vec{r}) \propto T(\vec{r})^\nu \text{ ,} \tag{1a}$$

where the polytropic index, determined by the exponent $\gamma$, or sometimes, by the exponent $\nu$, is

$$\gamma = 1 + 1/\nu \text{ or } \nu = 1/(\gamma - 1) \text{ .} \tag{1b}$$

Many space plasmas exhibit positive correlations between density and temperature, with their most frequent polytropic index close to the value of the adiabatic process, that is, $\gamma = 5/3$. Some examples are: solar wind proton and electron plasma (e.g., Totten et al. 1995; Newbury et al. 1997; Nicolaou et al. 2014a; Livadiotis & Desai 2016; Livadiotis 2018a), solar flares (e.g., Garcia 2001; Wang et al. 2015; 2016), planetary bow shocks (Winterhalter et al. 1984; Tatrallyay et al. 1984).

In several cases, space plasmas have polytropic indices with sub-adiabatic ($1 < \gamma < 5/3$) or super-adiabatic values ($5/3 < \gamma < +\infty$); for example in CMEs (e.g., Liu et al. 2006; Mishra & Wang 2018), coronal plasma (e.g., Prasad et al. 2018), Earth's plasma sheet (e.g., Zhu 1990; Goertz & Baumjohann 1991; Borovsky et al. 1998), planetary magnetospheres (e.g., Dialynas et al. 2018), even in the galaxy clusters and superclusters (e.g., Markevitch et al. 1998; Ettori et al. 2000; Grandi & Molendi 2002; Bautz et al. 2009).

However, there are rarer cases, where space plasmas have negative correlations between their density and temperature; these were found in the outer heliosphere (Elliott et al. 2019), inner heliosheath (e.g., Livadiotis et al. 2011; Livadiotis & McComas 2013), and the planetary magnetospheres, namely, terrestrial low latitude boundary layer (e.g., Sckopke et al. 1981), central plasma sheet (e.g., Pang et al. 2015), and bow shock (e.g., Pang et al. 2020); Jovian ionosphere (Allegrini et al. 2020), magnetosheath and boundary layer (e.g., Nicolaou et al. 2014b; 2015); and Saturnian magnetosphere (e.g., Dialynas et al. 2018).

Adiabatic processes characterize plasma flows with nearly zero heat transition; such example is the solar wind plasma particles that flow throughout the supersonic heliosphere under expansive cooling. However, turbulent heating disturbs this hypothetically ideal adiabatic process. The turbulence in solar wind has two sources: (i) the solar-origin large-scale energy fluctuations, and (ii) the excitation of plasma waves by newborn interstellar pickup ions. Other secondary sources may exist, e.g., turbulence can also be generated in-situ in the solar wind by velocity shears (e.g., Roberts et al. 1992, Zank et al. 2017). The turbulence affects the polytropic behavior of the solar wind plasma.

On the other hand, non-adiabatic polytropic indices can be connected with space plasmas residing in thermodynamic stationary states, namely, their particle distribution of velocity does not vary significantly with time. Particles of these plasmas are typically described by kappa distributions (Livadiotis & McComas 2010; 2013, Livadiotis 2017; 2018b); in particular, it has been theoretically shown that there is an equivalence between the polytropic behavior and the formalism of kappa distributions (e.g., Livadiotis 2019b).

Both the general cases of (i) adiabatic plasma flows, and (ii) non-adiabatic plasmas consistent with the formalism of kappa distributions, can be connected with temperature anisotropy caused by the presence of interplanetary / magnetospheric magnetic field. In general, space plasmas are described by anisotropic velocity distributions. Namely, the perpendicular and parallel directions, with respect to the magnetic field, have different thermal properties, while the respective temperature-like parameters are noted by $T_\perp$ and $T_\parallel$, and the anisotropy is defined by the ratio $\alpha \equiv T_\perp / T_\parallel$.

Up to this moment, there is no known theoretical connection between the polytropic index and the temperature anisotropy. Surely, theoretical models exist separately for the description of polytropic behavior (e.g., Livadiotis 2019b) and temperature anisotropy (e.g., Ao et al. 2003) in plasmas, but without any connection between them. For instance, histograms of the measurements of the polytropic indices and anisotropies for the solar wind proton plasma near 1 AU, computed using datasets taken from Wind S/C (e.g., Livadiotis & Desai 2016, Livadiotis 2018a, Nicolaou & Livadiotis 2019, Nicolaou et al. 2019), are shown in Figure 1. The polytropic indices are distributed with a mode near the adiabatic index, $\gamma \sim 1.67$, while the anisotropies are distributed with a mode near $\alpha \sim 0.8$. Then, we may ask: Is there a connection between the adiabatic polytropic indices included in this histogram and the values of temperature anisotropies? Is there a theoretical basis of such a relationship?

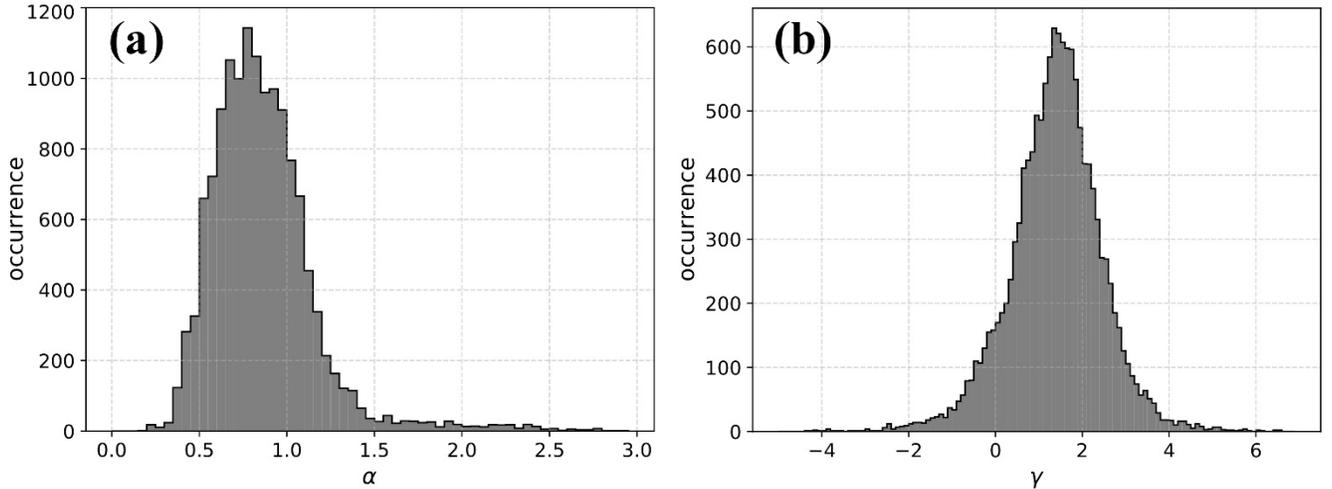

**Figure 1.** Histograms of the (a) temperature anisotropies, and (b) polytropic indices, for the solar wind proton plasma near 1 AU, computed using datasets taken from Wind S/C during the first 73 days of 1995. (Datasets will be described in §3.1, but they can be also found in Livadiotis & Desai 2016.)

The purpose of this paper is to (i) develop a theoretical model that connects the adiabatic polytropic index and the temperature anisotropy; a similar connection will be developed for the non-adiabatic polytropic indices; and (ii) perform a data and statistical analysis in order to examine the validity of the theoretically developed model in the case of the solar wind proton plasma near 1 AU; this will be performed using datasets taken from Wind S/C.

Section 2 presents the formalism of anisotropic kappa distributions and particle correlation, that is, the basic models of anisotropic kappa distributions, and how the derivation of the correlation coefficient; the latter will lead to the theoretically developed relationship between the adiabatic polytropic index and the anisotropy. Section 3 presents the application of this developed relationship to solar wind proton plasma near 1 au. We use plasma bulk parameters and interplanetary magnetic field datasets taken from Wind S/C, perform a data and statistical analysis to compare the observations with the developed relationship. In Section 4, we discuss the relationship of non-adiabatic polytropic indices with the anisotropy, which is the case for plasma flows under a dynamical potential; we also discuss related theoretical developments for future analyses, e.g., finding similar results using other models of anisotropic kappa distributions that may be less frequently or rarely used in space plasmas. Finally, conclusions summarize the results.

## 2. Formalism of anisotropic kappa distributions and particle correlation

### 2.1. Basic model of anisotropic kappa distributions

The typical 3D anisotropic kappa distribution is given by

$$P(u_\parallel, u_\perp; \theta_\parallel, \theta_\perp, \kappa) = [\pi(\kappa - \tfrac{3}{2})]^{-\frac{3}{2}} \cdot \frac{\Gamma(\kappa+1)}{\Gamma(\kappa - \tfrac{1}{2})} \cdot \theta_\parallel^{-1} \theta_\perp^{-2} \cdot \left[ 1 + \frac{1}{\kappa - \tfrac{3}{2}} \cdot \left( \frac{u_\parallel^2}{\theta_\parallel^2} + \frac{u_\perp^2}{\theta_\perp^2} \right) \right]^{-\kappa-1}, \qquad (2a)$$

with normalization

$$\int_0^\infty \int_{-\infty}^{+\infty} P(u_\parallel, u_\perp; \theta_\parallel, \theta_\perp, \kappa) \, du_\parallel \, 2\pi u_\perp du_\perp = 1 \; . \qquad (2b)$$

For applications of this model, see, e.g., Summers & Thorne 1991, Summers et al. 1994, Štverák et al. 2008, Astfalk et al. 2015, Khokhar et al. 2017, Wilson et al. 2019; Liu & Chen 2019; Khan et al. 2020). The velocity components are defined in correspondence to the direction of the magnetic field, that is, $\vec{u}_\parallel$ and $\vec{u}_\perp$ are set to be parallel and perpendicular to the field, respectively. The bulk velocities are taken into zero; for simplicity we consider the co-moving reference frame. The thermal speed of a particle of mass $m$, that is, $\theta = \sqrt{2k_\mathrm{B}T/m}$, denotes the temperature $T$ expressed in speed dimensions; $\theta^2/2$ provides the second statistical moment of the velocities (in the co-moving reference frame).

In order to calculate the correlation coefficient we need the formulation of multi-dimensional kappa distributions, that is, for describing a distribution of $N$ particles, where each particle is $d$-dimensional, (Livadiotis & McComas 2011; Livadiotis 2015a). Let the velocity vector of the $i$-th particle be $\vec{u}_i^2 = u_{i,x}^2 + u_{i,y}^2 + u_{i,z}^2$. In a $d$-dimensional velocity space, this is $\vec{u}_i^2 = u_{i,1}^2 + u_{i,2}^2 + \cdots + u_{i,d}^2$; therefore, the general case of a $N$-particle $N \cdot d$-dimensional kappa distribution is

$$P(\{\vec{u}_i\}_{i=1}^N; \theta, \kappa_0) = (\pi \kappa_0)^{-\frac{1}{2}Nd} \cdot \frac{\Gamma(\kappa_0 + 1 + \tfrac{1}{2}Nd)}{\Gamma(\kappa_0 + 1)} \cdot \theta^{-Nd} \cdot \left[ 1 + \frac{1}{\kappa_0} \cdot \frac{\vec{u}_1^2 + \vec{u}_2^2 + \cdots + \vec{u}_N^2}{\theta^2} \right]^{-\kappa_0 - 1 - \frac{1}{2}Nd}, \qquad (3a)$$

or in the specific case of $N$ 3D-particles,

$$P(\{\vec{u}_i\}_{i=1}^N; \theta, \kappa_0) = (\pi \kappa_0)^{-\frac{1}{2}Nd} \cdot \frac{\Gamma(\kappa_0 + 1 + \tfrac{3}{2}N)}{\Gamma(\kappa_0 + 1)} \cdot \theta^{-3N} \cdot \left[ 1 + \frac{1}{\kappa_0} \cdot \frac{\vec{u}_1^2 + \vec{u}_2^2 + \cdots + \vec{u}_N^2}{\theta^2} \right]^{-\kappa_0 - 1 - \frac{3}{2}N}, \qquad (3b)$$

where we denote $\{\vec{u}_i\}_{i=1}^N = \vec{u}_1, \vec{u}_2, \cdots, \vec{u}_N$. The involved kappa index $\kappa_0$ is independent on the dimensionality or the number of correlated particles involved in the kappa distribution. The standard kappa index $\kappa$ depends on the dimensionality, i.e., $\kappa(d) = const. + \tfrac{1}{2}d$. The involved constant is noted by $\kappa_0$, and indicates an invariant expression of the kappa index; thus, the kappa index that remains invariant under variations of the dimensionality or the number of the correlated degrees of freedom. We typically use the invariant kappa index, in order to express the kappa distributions of higher dimensionality.

The anisotropic kappa distribution shown in Eq.(2a) is a one-particle distribution that refers to systems with homogeneous correlation among the particle velocity components. The respective $N$-particle kappa distribution is given by

$$P\left(\{\vec{u}_i\}_{i=1}^N; \theta_\parallel, \theta_\perp, \kappa_0; N\right) = (\pi\,\kappa_0)^{-\frac{3}{2}N} \cdot \frac{\Gamma(\kappa_0 + 1 + \frac{3}{2}N)}{\Gamma(\kappa_0 + 1)} \cdot \theta_\parallel^{-N} \theta_\perp^{-2N}$$

$$\times \left[1 + \frac{1}{\kappa_0} \cdot \left(\frac{1}{\theta_\parallel^2}\sum_{i=1}^N u_{\parallel i}^2 + \frac{1}{\theta_\perp^2}\sum_{i=1}^N u_{\perp i}^2\right)\right]^{-\kappa_0 - 1 - \frac{3}{2}N} ,$$

(4a)

with normalization

$$\int_{-\infty}^{+\infty} \cdots (3N \text{ integrals}) \cdots \int_{-\infty}^{+\infty} P\left(\{\vec{u}_i\}_{i=1}^N; \theta_\parallel, \theta_\perp, \kappa_0; N\right) d\vec{u}_1 \ldots d\vec{u}_N = 1 ,$$

(4b)

where $d\vec{u}_i \equiv du_{\parallel i} du_{\perp x_i} du_{\perp y_i}$ .

In order to derive the correlation coefficient, we need to find first the covariance between two particle energies. For this, we use the two-particle 3D distributions (e.g., Swaczyna et al. 2019), that is, the isotropic distribution

$$P(\vec{u}_1, \vec{u}_2; \theta, \kappa_0) = (\pi\,\kappa_0)^{-3} \cdot \frac{\Gamma(\kappa_0 + 4)}{\Gamma(\kappa_0 + 1)} \cdot \theta^{-6} \cdot \left(1 + \frac{1}{\kappa_0} \cdot \frac{\vec{u}_1^2 + \vec{u}_2^2}{\theta^2}\right)^{-\kappa_0 - 4} ,$$

(5a)

while the respective anisotropic distribution is expressed as

$$P\left(\vec{u}_1, \vec{u}_2; \theta_\parallel, \theta_\perp, \kappa_0\right) = (\pi\,\kappa_0)^{-3} \cdot \frac{\Gamma(\kappa_0 + 4)}{\Gamma(\kappa_0 + 1)} \cdot \theta_\parallel^{-2} \theta_\perp^{-4}$$

$$\times \left[1 + \frac{1}{\kappa_0} \cdot \left(\frac{u_{\parallel 1}^2 + u_{\parallel 2}^2}{\theta_\parallel^2} + \frac{u_{\perp 1}^2 + u_{\perp 2}^2}{\theta_\perp^2}\right)\right]^{-\kappa_0 - 4} .$$

(5b)

In terms of the standard 3D kappa index $\kappa$, the two-particle isotropic and anisotropic 3D distributions are respectively given by:

$$P(\vec{u}_1, \vec{u}_2; \theta, \kappa) = [\pi(\kappa - \tfrac{3}{2})]^{-3} \cdot \frac{\Gamma(\kappa + \frac{5}{2})}{\Gamma(\kappa - \frac{1}{2})} \cdot \theta^{-6} \cdot \left(1 + \frac{1}{\kappa - \frac{3}{2}} \cdot \frac{\vec{u}_1^2 + \vec{u}_2^2}{\theta^2}\right)^{-\kappa - \frac{5}{2}} ,$$

(6a)

and

$$P\left(\vec{u}_1, \vec{u}_2; \theta_\parallel, \theta_\perp, \kappa_0\right) = [\pi(\kappa - \tfrac{3}{2})]^{-3} \cdot \frac{\Gamma(\kappa + \frac{5}{2})}{\Gamma(\kappa - \frac{1}{2})} \cdot \theta_\parallel^{-2} \theta_\perp^{-4}$$

$$\times \left[1 + \frac{1}{\kappa - \frac{3}{2}} \cdot \left(\frac{u_{\parallel 1}^2 + u_{\parallel 2}^2}{\theta_\parallel^2} + \frac{u_{\perp 1}^2 + u_{\perp 2}^2}{\theta_\perp^2}\right)\right]^{-\kappa - \frac{5}{2}} .$$

(6b)

## 2.2. Correlation coefficient

Here we present the derivation of the correlation coefficient that characterizes the particle energies in a population described by anisotropic kappa distributions. The Pearson's correlation coefficient (Abe 1999; Livadiotis & McComas 2011; Livadiotis 2015a; 2017, Ch.5.4; Livadiotis et al. 2020) is given by

$$\rho = \frac{\sigma_{\varepsilon_1 \varepsilon_2}^2}{\sigma_{\varepsilon\varepsilon}^2} = \frac{\sigma_{u_1^2 u_2^2}^2}{\sigma_{u^2 u^2}^2} ,$$

(7a)

where

$$\sigma_{\varepsilon_1\varepsilon_2}^2 = \langle \varepsilon_1\varepsilon_2 \rangle - \langle \varepsilon \rangle^2 = \tfrac{1}{2}m\left( \langle u_1^2 u_2^2 \rangle - \langle u^2 \rangle^2 \right) = \tfrac{1}{2}m\sigma_{u_1^2 u_2^2}^2 \ , \tag{7b}$$

and

$$\sigma_{\varepsilon\varepsilon}^2 = \langle \varepsilon^2 \rangle - \langle \varepsilon \rangle^2 = \tfrac{1}{2}m\left( \langle u^4 \rangle - \langle u^2 \rangle^2 \right) = \tfrac{1}{2}m\sigma_{u^2 u^2}^2 \ , \tag{7c}$$

Using the isotropic kappa distributions (Livadiotis & McComas 2011), the correlation coefficient was found:

$$\rho = \frac{\tfrac{1}{2}d}{\kappa_0 + \tfrac{1}{2}d} \ \text{ or } \ \rho = \frac{\tfrac{1}{2}d}{(\kappa - \tfrac{3}{2}) + \tfrac{1}{2}d} \ , \tag{8a}$$

or, in the case of 3D distributions,

$$\rho = \frac{\tfrac{3}{2}}{\kappa_0 + \tfrac{3}{2}} \ \text{ or } \ \rho = \frac{\tfrac{3}{2}}{\kappa} \ . \tag{8b}$$

It is important to note that while the mean particle kinetic energy provides the kinetic definitions of thermal energy, $k_B T$, the (normalized) covariance (that is, the correlation coefficient) of the particle kinetic energy provides the kinetic definition of the (inverse) kappa index, $1/\kappa$.

In the case of the anisotropic kappa distributions, we find the variance

$$\sigma_{u^2 u^2}^2 = \langle u^4 \rangle - \langle u^2 \rangle^2 = (\tfrac{1}{2}\theta_\parallel^4 + \theta_\perp^4) + (\tfrac{3}{4}\theta_\parallel^4 + 2\theta_\perp^4 + \theta_\parallel^2\theta_\perp^2)\frac{1}{\kappa - \tfrac{5}{2}} \ , \tag{9a}$$

and covariance

$$\sigma_{u_1^2 u_2^2}^2 = \langle u_1^2 u_2^2 \rangle - \langle u^2 \rangle^2 = (\tfrac{1}{4}\theta_\parallel^4 + \theta_\perp^4 + \theta_\parallel^2\theta_\perp^2) \cdot \frac{1}{\kappa - \tfrac{5}{2}} \ . \tag{9b}$$

Substituting Eqs.(9a,b) in Eq.(7a), we obtain

$$\rho = \frac{\sigma_{u_1^2 u_2^2}^2}{\sigma_{u^2 u^2}^2} = \frac{\dfrac{\tfrac{1}{4}\theta_\parallel^4 + \theta_\perp^4 + \theta_\parallel^2\theta_\perp^2}{\tfrac{1}{2}\theta_\parallel^4 + \theta_\perp^4}}{(\kappa - \tfrac{3}{2}) + \dfrac{\tfrac{1}{4}\theta_\parallel^4 + \theta_\perp^4 + \theta_\parallel^2\theta_\perp^2}{\tfrac{1}{2}\theta_\parallel^4 + \theta_\perp^4}} \ . \tag{10}$$

Introducing the temperature anisotropy by

$$\alpha \equiv T_\perp / T_\parallel \ , \tag{11}$$

and given that the temperature can be written as $T = \tfrac{1}{3}(T_\parallel + 2T_\perp)$, then, the temperature-like components are expressed in terms of the actual temperature and anisotropy as

$$T_\parallel = \frac{3}{1 + 2\alpha} \cdot T \ , \ T_\perp = \frac{3\alpha}{1 + 2\alpha} \cdot T \ . \tag{12a}$$

In terms of thermal speeds, Eq.(12a) becomes

$$\theta_\parallel^2 = \frac{3}{1 + 2\alpha} \cdot \theta^2 \ , \ \theta_\perp^2 = \frac{3\alpha}{1 + 2\alpha} \cdot \theta^2 \ . \tag{12b}$$

Then, the correlation coefficient in Eq.(10) is written as

$$\rho = \frac{\frac{1}{2}\frac{(2\alpha+1)^2}{2\alpha^2+1}}{(\kappa-\frac{3}{2})+\frac{1}{2}\frac{(2\alpha+1)^2}{2\alpha^2+1}} \quad . \tag{13}$$

Then, the correlation coefficient can be expressed in terms of an effective dimensionality, i.e.,

$$\rho = \frac{\frac{1}{2}d_{\text{eff}}}{\kappa_0+\frac{1}{2}d_{\text{eff}}} \text{ or } \rho = \frac{\frac{1}{2}d_{\text{eff}}}{(\kappa-\frac{3}{2})+\frac{1}{2}d_{\text{eff}}} \quad . \tag{14}$$

The effective dimensionality recovers the dimensionality of the embedded 3D velocity space in the isotropic case $\alpha=1$. In general, the impact of anisotropy on dimensionality is non trivial: The dimensionality of the embedded space is $d=3-$ for any anisotropy; however, the limiting cases where the parallel or perpendicular directions are neglected should characterize a degeneration of the velocity distributions so that its dimensionality to be reduced to 2D or 1D, respectively. Then, the effective dimensionality defined with respect to anisotropy, $d_{\text{eff}}$, is expected to satisfy:

$$\begin{aligned} \alpha &= 0 \quad \Leftrightarrow \quad d_{\text{eff}} = 2, \\ \alpha &= 1 \quad \Leftrightarrow \quad d_{\text{eff}} = 3, \\ \alpha &= \infty \quad \Leftrightarrow \quad d_{\text{eff}} = 1. \end{aligned} \tag{15}$$

The relationship between the effective dimensionality and anisotropy is derived from comparing the correlation coefficient in Eq.(13) and Eq. (14), i.e.,

$$d_{\text{eff}}(\alpha) = \frac{(2\alpha+1)^2}{2\alpha^2+1} \quad , \tag{16}$$

where we observe that satisfies the extreme conditions of Eq.(15).

Next, we connect the expression in Eq.(16) with the polytropic index that describes adiabatic thermodynamic processes. The adiabatic polytropic index is given in terms of the effective kinetic degrees of freedom (e.g., Livadiotis 2015b) as

$$\gamma = 1 + 2d_{\text{eff}}^{-1} \quad , \tag{17}$$

therefore, we end up with the relationship

$$\gamma(\alpha) = 1 + \frac{\alpha^2+\frac{1}{2}}{(\alpha+\frac{1}{2})^2} \quad . \tag{18}$$

Figure 3 plots the effective dimensionality and the adiabatic polytropic index with respect to the anisotropy. For anisotropy $\alpha=1$, the distribution is spherical, and the effective dimensionality equals the space dimensionality, $d_{\text{eff}}=3$. As the anisotropy decreases for $\alpha<1$, reaching $\alpha=0$, the dimensionality decreases reaching $d_{\text{eff}}=1$, describing a linear distribution (like a cigar); as the anisotropy increases for $\alpha>1$, reaching $\alpha\to\infty$, the dimensionality decreases reaching $d_{\text{eff}}=2$, describing a "flat" distribution (like a pie). The effective dimensionality $d_{\text{eff}}$ has a local maximum at $\alpha=1$, while the corresponding adiabatic polytropic index $\gamma$ has a local minimum at $\alpha=1$.

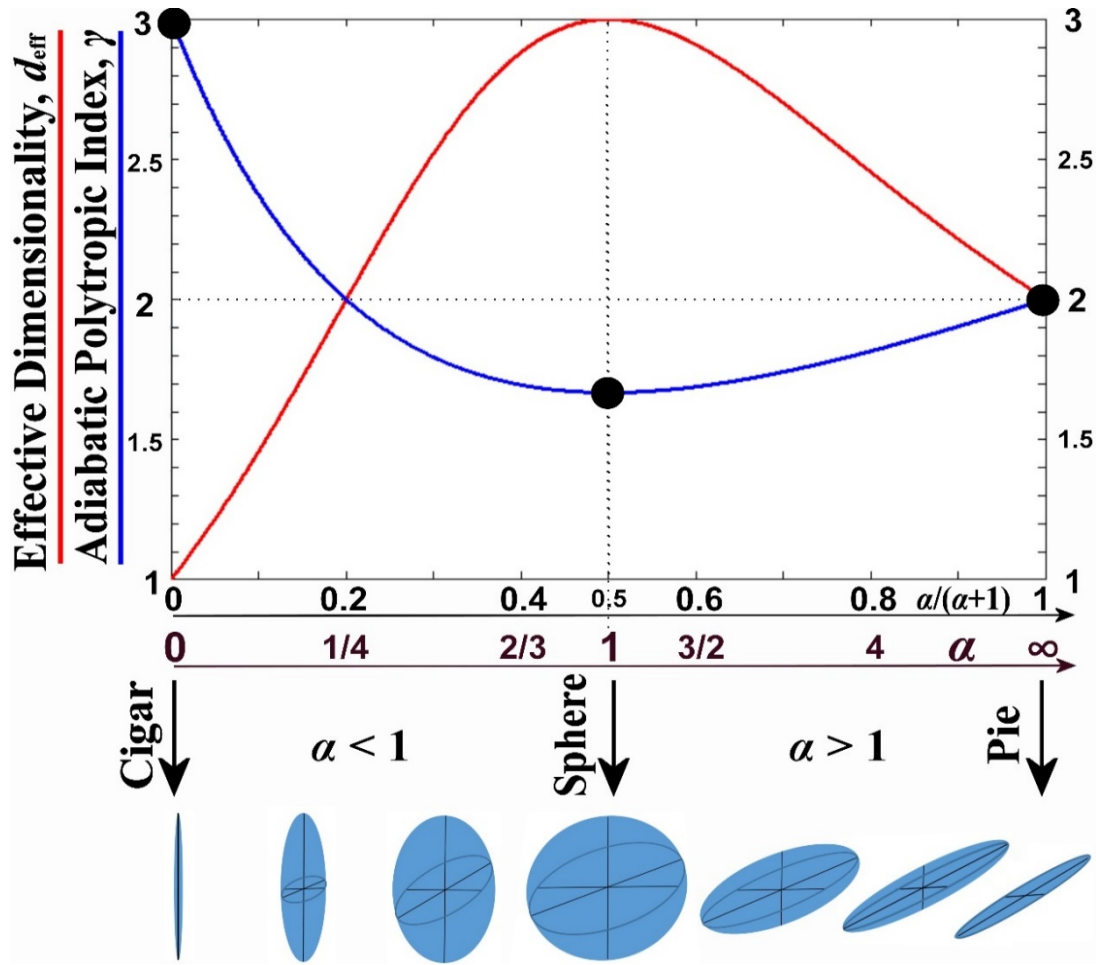

**Figure 2.** Effective dimensionality $d_{eff}$ (red) and the respective adiabatic polytropic index $\gamma$ (blue) as a function of the anisotropy $\alpha$, emphasizing the cases of $\alpha$=0, 1, $\infty$, corresponding to $d_{eff}$=1, 2, 3, and $\gamma$=3, 5/3, 2, where the velocity distribution takes the form of a cigar, sphere, pie. (Taken from Livadiotis et al. 2020)

### 3. Application to solar wind proton plasma near 1 au

*3.1. Data*

We show the dependence of the adiabatic polytropic index on the anisotropy of the velocity distribution for the solar wind proton plasma near 1 AU. We use publicly available data of solar wind proton bulk parameters and interplanetary magnetic field, taken from the Wind mission.

In particular, we use ~92-second solar wind plasma moments (speed $V_{sw}$, anisotropy $\alpha$, density $n$, and temperature $T$) (Ogilvie et al. 1995) and simultaneous measurements of the interplanetary magnetic field (Lepping et al. 1995), measured, respectively, from SWE and MFI instruments onboard Wind S/C, during the first 73 days of 1995 (see Figure 3). This time period occurred during the declining phase of solar activity cycle 23, and was characterized by corotating interaction regions that are apparent in increases in the solar wind density and magnetic field magnitude that precede the arrival of the high speed streams at 1 AU (e.g., Jian et al. 2006a;b; 2009).

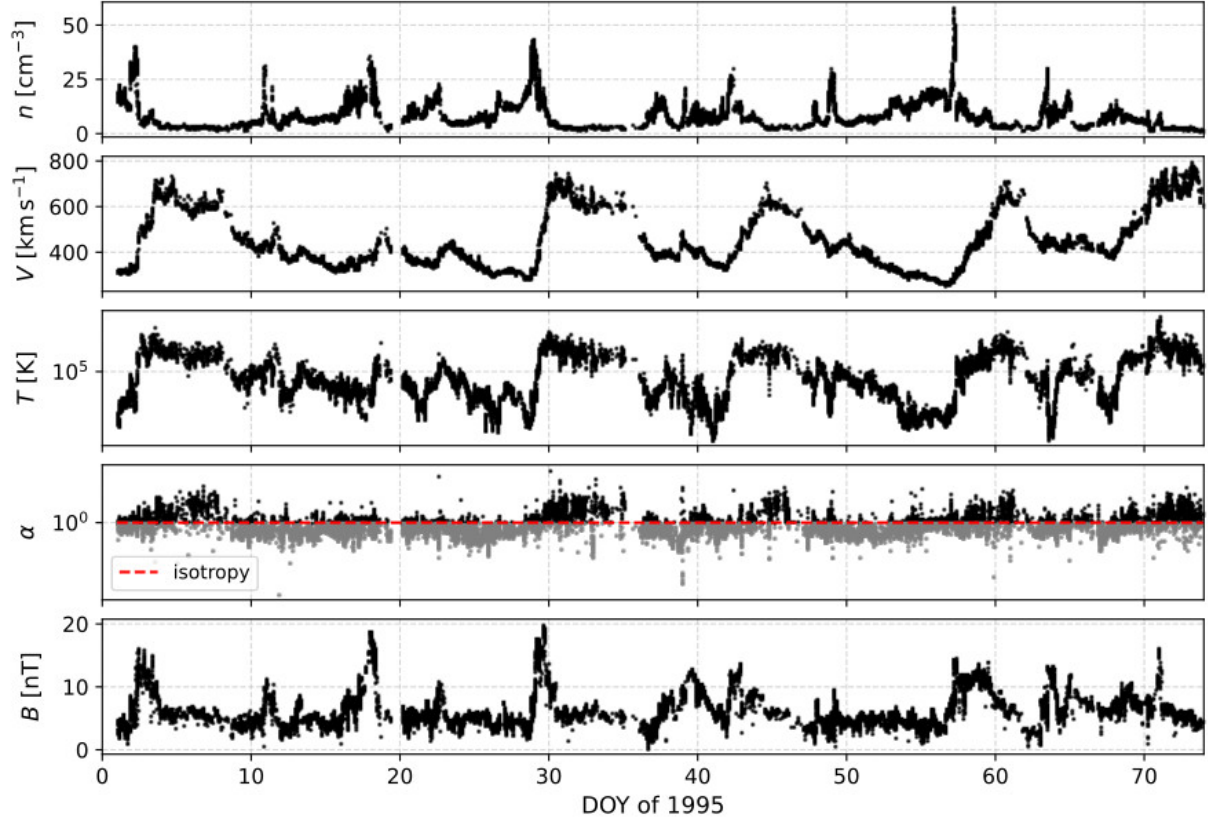

**Figure 3.** Datasets used in this study: ~92 s resolution measurements of the bulk solar wind plasma parameters and magnetic field magnitude, taken respectively from SWE and MFI instruments onboard Wind, during the first 73 days of 1995.

### 3.2. Method

We derive the polytropic index using the proton plasma moments and interplanetary magnetic field time series shown in Figure 3, according to the following steps:

- *Selection of time intervals*: Analyze the time intervals for which we fit $T$ and $n$. The intervals must be short enough to minimize the possibility to mix measurements of different streamlines, but large enough to improve statistics. Following previous studies (e.g., Newbury et al. 1997, Kartalev 2006; Nicolaou et al. 2014a), we select intervals covering 8 consecutive measurements of $T$ and $n$.

- *Filtering Bernoulli's integral*: For each interval, we examine the variability of Bernoulli's energy integral (Kartalev 2006). The standard deviation of the 8 values of the Bernoulli's integral must be <10% of their mean. The threshold was chosen to be low enough to minimize the possibility of having different streamlines in an interval, but also high enough to retain a statistically significant number of data points.

- *Fitting*: For each interval, we fit with a linear statistical model the logarithms of temperature and density (for the log-normal distributions of plasma moments, see: Burlaga & Lazarus 2000; Kasper et al. 2006), i.e., log $T$ vs. log $n$, where the slope equals $\gamma-1$. (See the example in Figure 4).

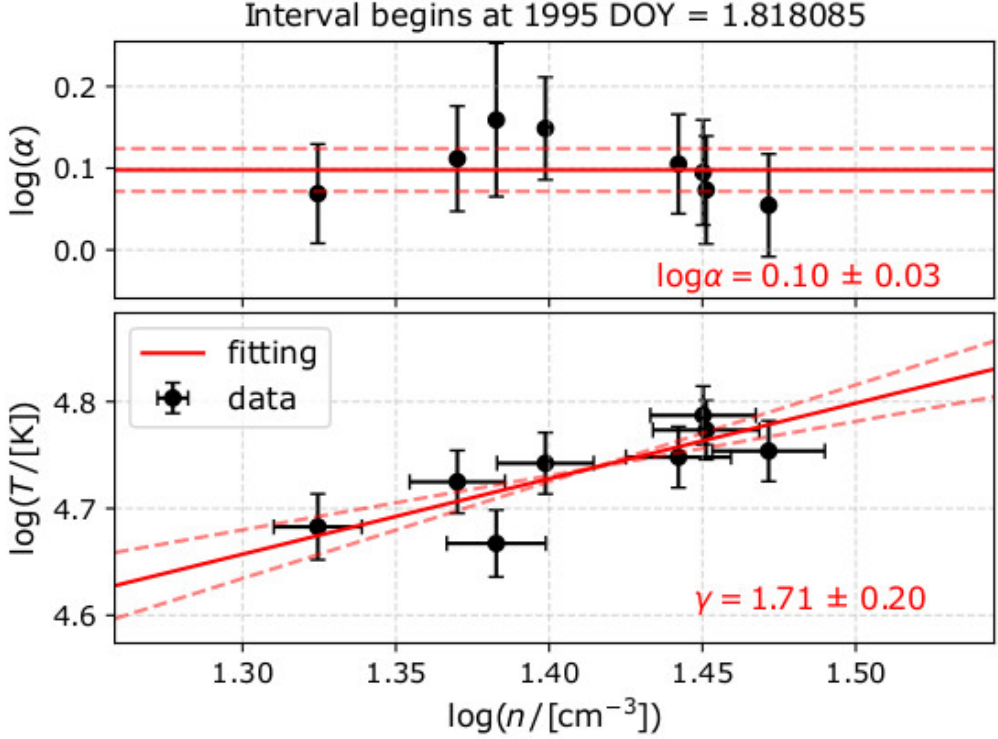

**Figure 4.** Estimation of the polytropic index (lower) and the average anisotropy (upper). The double-error weighted fitting of the temperature $T$ and density $n$ on log-log scales leads to the estimation of the slope ($\gamma$-1), and thus, to the polytropic index (in this example, we compute the anisotropy $\alpha$~1.3 and the polytropic index $\gamma$≈1.71).

### 3.3. Statistical analysis

Having the polytropic indices and their uncertainties derived from the linear fits for all the time intervals of the examined datasets (Figure 3), we analyze the constructed dataset of polytropic indices against the anisotropies. The 2D-histogram of the polytropic indices against the anisotropies is normalized by dividing it with the 1D-histogram of anisotropies; in particular, the number of data points at each $\Delta\alpha\times\Delta\gamma$ 2D-bin is divided by the number of data points in the whole column of the 1D $\Delta\alpha$-bin. (This type of standardized 2D histogram is called conditional 2D-histogram, corresponding to the ratio of a joint 2D probability $P(x,y)$ over $P(x)$; for more details, see: Livadiotis & Desai 2016, Park 2018, p.99.) The constructed normalized $\alpha\times\gamma$ 2D-histogram is plotted in the background of Figure 5. The co-plotted black-dash curve shows the modeled relationship between adiabatic polytropic indices and anisotropies as described by Eq.(18).

Next, we average all the polytropic indices corresponding in each column of the 1D $\Delta\alpha$-bin. The averaging is performed in two different ways. First, we estimate the weighted means and errors, each corresponding to each $\Delta\alpha$-bin, leading to the data points plotted in Figure 5(a). In another way, we use the normalized 2D-histogram that provides the probability of having a polytropic index at each $\Delta\alpha\times\Delta\gamma$ 2D-bin; in particular, we estimate the average by summing the occurrence given by the normalized 2D-histogram multiplied by the value of the polytropic index at the center of each bin; we count only the bins with a number of data points $N$>10, leading to the averaged data points plotted in Figure 5(b). Note that in both the averaging ways, we include only indices in the interval $0 \leq \gamma \leq 3$, corresponding to a reasonable set of values of dimensionality, that is, $1 \leq d_{\text{eff}} < \infty$; (Figure 5 demonstrates only the subintervals $1 \leq d_{\text{eff}} < 3$ and $1 \leq \gamma \leq 3$). The necessary formulae for the two ways are:

- For the weighted average and error:

$$\bar{\gamma} = \sum_{i=1}^{N} \delta\gamma_i^{-2} \gamma_i \Big/ \sum_{i=1}^{N} \delta\gamma_i^{-2} \quad , \quad \delta\bar{\gamma} = \sqrt{(\delta\bar{\gamma}^{prop})^2 + (\delta\bar{\gamma}^{stat})^2} \quad , \tag{19a}$$

where the involved two types of errors are the propagation and statistical ones, respectively:

$$\delta\bar{\gamma}^{prop} = \frac{1}{\sqrt{\sum_{i=1}^{N} \delta\gamma_i^{-2}}} \quad , \quad \delta\bar{\gamma}^{stat} = \sqrt{\frac{\frac{1}{N}\sum_{i=1}^{N} \delta\gamma_i^{-2}(\gamma_i - \bar{\gamma})^2}{\sum_{i=1}^{N} \delta\gamma_i^{-2} - \left(\sum_{i=1}^{N} \delta\gamma_i^{-2}\right)^{-1} \cdot \sum_{i=1}^{N} \delta\gamma_i^{-4}}} \quad , \tag{19b}$$

(see: Livadiotis 2007; 2018a).

- For the average derived from the normalized histogram $\{p_j\}$:

$$\bar{\gamma} = \sum_{j=1}^{M} p_j \gamma_j \quad , \quad \delta\bar{\gamma} = \sum_{j=1}^{M} p_j (\gamma_j - \bar{\gamma})^2 \quad . \tag{19c}$$

Then, we combine the two averaged polytropic indices at each column of the 1D $\Delta\alpha$-bin, using as a weight both the inverse variance and the reduced chi-square that measures the fit goodness, i.e.,

$$\bar{\gamma} = \sum_{i=1}^{2} w_i \delta\gamma_i^{-2} \gamma_i \Big/ \sum_{i=1}^{2} w_i \delta\gamma_i^{-2} \quad , \quad w_i = \left|1 - \chi_{\text{red},i}^2\right| \quad . \tag{19d}$$

The two datasets of averaged polytropic indices, noted by $\{\gamma_i^{(1)} \pm \delta\gamma_i^{(1)}\}_{i=1}^{N_0}$ and $\{\gamma_i^{(2)} \pm \delta\gamma_i^{(2)}\}_{i=1}^{N_0}$, are plotted in Figure 5(a) and (b), respectively, while their combined dataset, derived from weighted averaging of the two plots and noted by $\{\gamma_i^{(C_1)} \pm \delta\gamma_i^{(C_1)}\}_{i=1}^{N_0}$, is plotted in Figure 5(c). Below we describe an alternative way of averaging the two plots that mitigates potential bias caused by false statistical weighting; we note that this alternative averaging leads to the combined dataset $\{\gamma_i^{(C_2)} \pm \delta\gamma_i^{(C_2)}\}_{i=1}^{N_0}$ plotted in Figure 6. Moreover, Figure 5(d) plots a double-combined dataset, noted by $\{\gamma_i^{(CC)} \pm \delta\gamma_i^{(CC)}\}_{i=1}^{N_0}$, derived from both the weighted averages of $\{\gamma_i^{(C_1)} \pm \delta\gamma_i^{(C_1)}\}_{i=1}^{N_0}$ and $\{\gamma_i^{(C_2)} \pm \delta\gamma_i^{(C_2)}\}_{i=1}^{N_0}$.

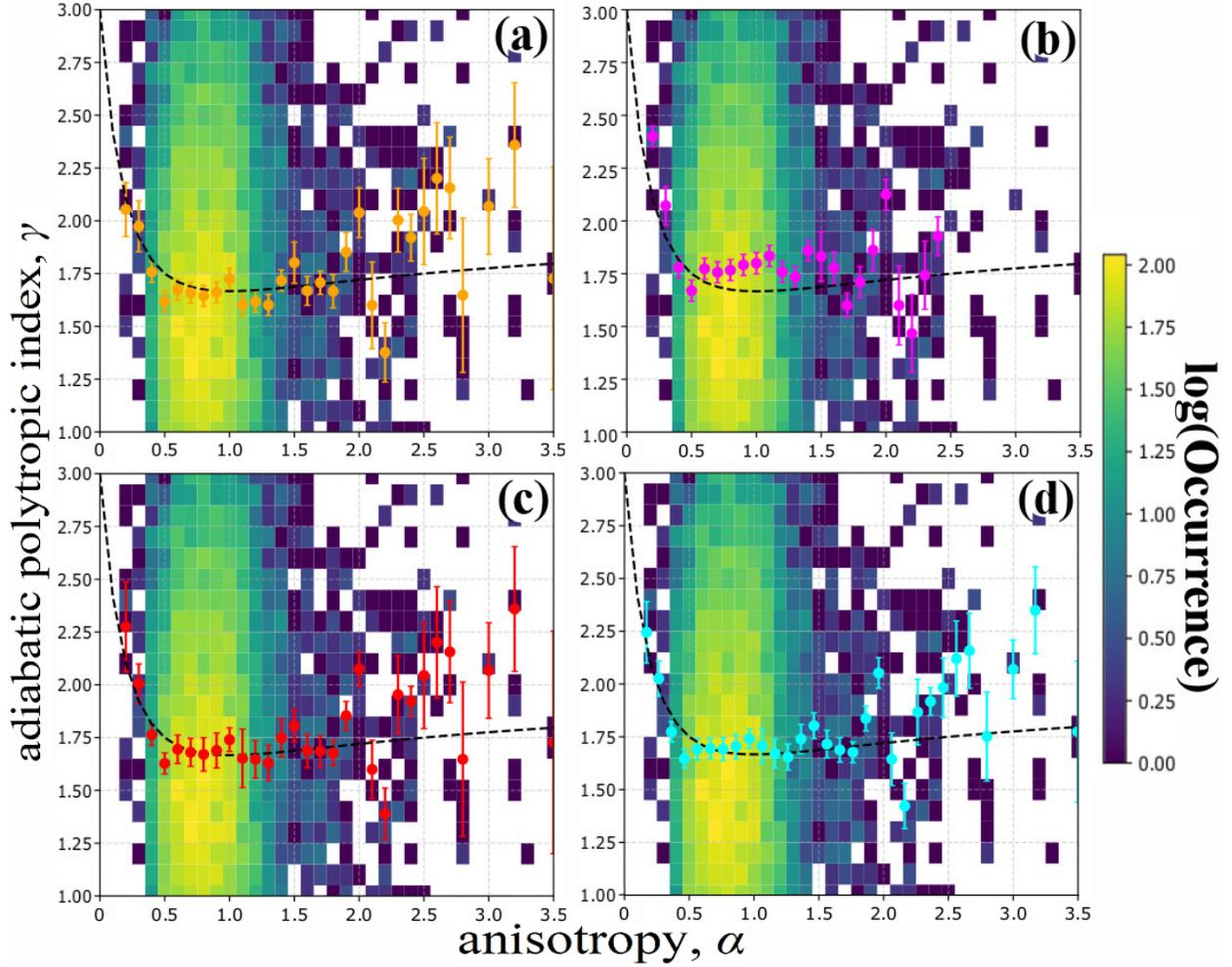

**Figure 5.** Plot of adiabatic polytropic indices vs. anisotropies. The normalized 2D-histogram and the modeled relationship of Eq.(39b) (black dash curve) are plotted in each panel; (note that the color-map measures the original occurrence on a log scale, that is, the logarithm of the pixel counts). Plotted datasets: (a) $\{\gamma_i^{(1)} \pm \delta\gamma_i^{(1)}\}_{i=1}^{N_0}$ (orange), derived from the weighted mean of polytropic indices at each $\Delta\alpha$-bin of the 1D-histogram of anisotropies; (b) $\{\gamma_i^{(2)} \pm \delta\gamma_i^{(2)}\}_{i=1}^{N_0}$ (magenta), derived from the normalized 2D-histogram; (c) $\{\gamma_i^{(C_1)} \pm \delta\gamma_i^{(C_1)}\}_{i=1}^{N_0}$ (red), combined from (a) and (b) datasets; and (d) $\{\gamma_i^{(CC)} \pm \delta\gamma_i^{(CC)}\}_{i=1}^{N_0}$ (light blue), combined from (c) and the homogenized dataset $\{\gamma_i^{(C_2)} \pm \delta\gamma_i^{(C_2)}\}_{i=1}^{N_0}$, plotted in Figure 6.

Next, we perform a specific statistical analysis that examines the fit stability and homogenization, and is appropriate for determining whether the fit of the modeled adiabatic polytropic indices to their averaged values, suffers from any bias of false statistical weighting caused by cofounding variables or other systematic errors. In particular, for each of the given set of $N_0$ data points of $\{\alpha_i \pm \delta\alpha_i, \gamma_i \pm \delta\gamma_i\}_{i=1}^{N_0}$, we (a) reproduce $N$=1000 bi-normally distributed data points, $\sim \mathcal{N}\{\mu_x \equiv \alpha_i, \sigma_x \equiv \delta\alpha_i\} \times \mathcal{N}\{\mu_y \equiv \gamma_i, \sigma_y \equiv \delta\gamma_i\}$; (for further details, see: Livadiotis 2016); (b) homogenize the enriched dataset of $N \times N_0$ data points by averaging it over the original 1D $\Delta\alpha$-binning, leading to the new dataset $\{\gamma_i^{(C_2)} \pm \delta\gamma_i^{(C_2)}\}_{i=1}^{N_0}$; (c) examine the fitting of the derived averaged indices $\{\gamma_i^{(C_2)} \pm \delta\gamma_i^{(C_2)}\}_{i=1}^{N_0}$ with the modeled relationship between the adiabatic polytropic indices and the anisotropies; this should lead to the same statistical results as the original dataset $\{\alpha_i \pm \delta\alpha_i, \gamma_i \pm \delta\gamma_i\}_{i=1}^{N_0}$, otherwise, the presence of potential cofounding variables would have led to statistically significant difference.

We have performed an overall averaging between the two combined datasets, namely, those combined by (i) $\{\gamma_i^{(C_1)} \pm \delta\gamma_i^{(C_1)}\}_{i=1}^{N_0}$, i.e., the direct averaging as plotted in Figure 5(c), and (ii) $\{\gamma_i^{(C_2)} \pm \delta\gamma_i^{(C_2)}\}_{i=1}^{N_0}$, i.e., the homogenization of the enriched dataset as plotted in Figure 6; the resulted dataset, $\{\gamma_i^{(CC)} \pm \delta\gamma_i^{(CC)}\}_{i=1}^{N_0}$, is shown in Figure 5(d). (It is noted that the method of homogenization of data is required when deriving polytropic indices; e.g., misleading measurements of polytropic indices due to inhomogeneities in the solar wind originating at coronal source regions have been detected and mitigated by Newbury et al. 1997.)

### 3.4. Results

Figure 7 shows the 2D-histogram of the enriched $N \times N_0$ dataset of $\gamma$ vs. $\alpha$ values, the 1D histograms of $\alpha$ and $\gamma$, and the homogenized set of polytropic indices $\{\gamma_i^{(C_2)} \pm \delta\gamma_i^{(C_2)}\}_{i=1}^{N_0}$, together with the modeled relationship in Eq.(18). The $p$-value of the extremes is a measure of the goodness of fits. The fitting of the model to the optimized dataset $\{\gamma_i^{(C_2)} \pm \delta\gamma_i^{(C_2)}\}_{i=1}^{N_0}$ is characterized by $p\sim0.082$ ($>0.05$), corresponding to a statistically confident agreement of the model with the observed dataset. All the results of the statistical analysis are shown in Table 1.

The derived theoretical model $\gamma(\alpha)$ has zero flexible parameters to be fitted; yet, the high $p$-value indicating the goodness of the fitting and its high statistical confidence.

**Table 1.** Results of the statistical analysis of adiabatic polytropic indices per anisotropy

| (1) | | | (2) | | | (C₁) | | | (C₂) | | | | (CC) | | | |
|---|---|---|---|---|---|---|---|---|---|---|---|---|---|---|---|---|
| $\alpha$ | $\gamma$ | $\delta\gamma$ | $\alpha$ | $\gamma$ | $\delta\gamma$ | $\alpha$ | $\gamma$ | $\delta\gamma$ | $\alpha$ | $\delta\alpha$ | $\gamma$ | $\delta\gamma$ | $\alpha$ | $\delta\alpha$ | $\gamma$ | $\delta\gamma$ |
| 0.2 | 2.05 | 0.13 | 0.2 | 2.40 | 0.05 | 0.2 | 2.27 | 0.21 | 0.16 | 0.03 | 2.22 | 0.20 | 0.17 | 0.03 | 2.24 | 0.15 |
| 0.3 | 1.97 | 0.12 | 0.3 | 2.07 | 0.09 | 0.3 | 2.01 | 0.09 | 0.25 | 0.03 | 2.12 | 0.20 | 0.26 | 0.03 | 2.03 | 0.08 |
| 0.4 | 1.76 | 0.05 | 0.4 | 1.78 | 0.05 | 0.4 | 1.76 | 0.05 | 0.35 | 0.03 | 1.90 | 0.17 | 0.36 | 0.02 | 1.77 | 0.05 |
| 0.5 | 1.62 | 0.05 | 0.5 | 1.67 | 0.05 | 0.5 | 1.63 | 0.05 | 0.45 | 0.03 | 1.71 | 0.10 | 0.46 | 0.03 | 1.65 | 0.04 |
| 0.6 | 1.67 | 0.05 | 0.6 | 1.77 | 0.05 | 0.6 | 1.69 | 0.07 | 0.55 | 0.03 | 1.69 | 0.08 | 0.56 | 0.03 | 1.69 | 0.05 |
| 0.7 | 1.66 | 0.05 | 0.7 | 1.76 | 0.05 | 0.7 | 1.68 | 0.07 | 0.65 | 0.03 | 1.71 | 0.07 | 0.66 | 0.03 | 1.69 | 0.05 |
| 0.8 | 1.64 | 0.05 | 0.8 | 1.77 | 0.05 | 0.8 | 1.67 | 0.08 | 0.75 | 0.03 | 1.71 | 0.07 | 0.76 | 0.03 | 1.69 | 0.05 |
| 0.9 | 1.66 | 0.05 | 0.9 | 1.79 | 0.05 | 0.9 | 1.69 | 0.08 | 0.85 | 0.03 | 1.72 | 0.08 | 0.86 | 0.02 | 1.70 | 0.06 |
| 1.0 | 1.72 | 0.05 | 1.0 | 1.80 | 0.05 | 1.0 | 1.74 | 0.06 | 0.95 | 0.03 | 1.74 | 0.08 | 0.96 | 0.03 | 1.74 | 0.04 |
| 1.1 | 1.60 | 0.05 | 1.1 | 1.83 | 0.05 | 1.1 | 1.65 | 0.14 | 1.05 | 0.03 | 1.74 | 0.10 | 1.06 | 0.02 | 1.71 | 0.08 |
| 1.2 | 1.62 | 0.05 | 1.2 | 1.76 | 0.05 | 1.2 | 1.65 | 0.09 | 1.15 | 0.03 | 1.71 | 0.11 | 1.16 | 0.02 | 1.67 | 0.07 |
| 1.3 | 1.60 | 0.05 | 1.3 | 1.73 | 0.05 | 1.3 | 1.63 | 0.08 | 1.25 | 0.03 | 1.68 | 0.09 | 1.26 | 0.02 | 1.65 | 0.06 |
| 1.4 | 1.72 | 0.05 | 1.4 | 1.86 | 0.05 | 1.4 | 1.75 | 0.09 | 1.35 | 0.03 | 1.73 | 0.11 | 1.36 | 0.03 | 1.74 | 0.07 |
| 1.5 | 1.80 | 0.10 | 1.5 | 1.83 | 0.12 | 1.5 | 1.81 | 0.08 | 1.45 | 0.03 | 1.80 | 0.10 | 1.46 | 0.03 | 1.80 | 0.06 |
| 1.6 | 1.67 | 0.07 | 1.6 | 1.78 | 0.08 | 1.6 | 1.69 | 0.08 | 1.55 | 0.03 | 1.77 | 0.11 | 1.56 | 0.03 | 1.71 | 0.07 |
| 1.7 | 1.71 | 0.06 | 1.7 | 1.60 | 0.06 | 1.7 | 1.69 | 0.07 | 1.65 | 0.03 | 1.69 | 0.09 | 1.66 | 0.02 | 1.69 | 0.06 |
| 1.8 | 1.67 | 0.08 | 1.8 | 1.71 | 0.07 | 1.8 | 1.68 | 0.06 | 1.75 | 0.03 | 1.68 | 0.09 | 1.76 | 0.03 | 1.68 | 0.05 |
| 1.9 | 1.85 | 0.09 | 1.9 | 1.86 | 0.10 | 1.9 | 1.85 | 0.07 | 1.85 | 0.03 | 1.78 | 0.13 | 1.86 | 0.03 | 1.84 | 0.06 |

| | | | | | | | | | | | | | | | | | |
|---|---|---|---|---|---|---|---|---|---|---|---|---|---|---|---|---|---|
| 2.0 | 2.04 | 0.12 | 2.0 | 2.13 | 0.07 | 2.0 | 2.07 | 0.08 | 1.95 | 0.03 | 1.96 | 0.16 | 1.96 | 0.02 | 2.05 | 0.07 |
| 2.1 | 1.60 | 0.21 | 2.1 | 1.60 | 0.19 | 2.1 | 1.60 | 0.14 | 2.05 | 0.03 | 1.83 | 0.29 | 2.06 | 0.03 | 1.64 | 0.12 |
| 2.2 | 1.38 | 0.14 | 2.2 | 1.47 | 0.18 | 2.2 | 1.39 | 0.12 | 2.15 | 0.03 | 1.53 | 0.22 | 2.16 | 0.03 | 1.42 | 0.11 |
| 2.3 | 2.00 | 0.15 | 2.3 | 1.74 | 0.16 | 2.3 | 1.95 | 0.19 | 2.25 | 0.03 | 1.66 | 0.29 | 2.26 | 0.03 | 1.87 | 0.16 |
| 2.4 | 1.92 | 0.11 | 2.4 | 1.93 | 0.09 | 2.4 | 1.92 | 0.07 | 2.35 | 0.03 | 1.88 | 0.18 | 2.36 | 0.02 | 1.92 | 0.07 |
| 2.5 | 2.04 | 0.25 | 2.5 | - | - | 2.5 | 2.04 | 0.25 | 2.45 | 0.03 | 1.95 | 0.17 | 2.46 | 0.02 | 1.98 | 0.14 |
| 2.6 | 2.20 | 0.26 | 2.6 | - | - | 2.6 | 2.20 | 0.26 | 2.55 | 0.03 | 2.05 | 0.24 | 2.56 | 0.03 | 2.12 | 0.18 |
| 2.7 | 2.16 | 0.24 | 2.7 | - | - | 2.7 | 2.16 | 0.24 | 2.65 | 0.03 | 2.16 | 0.26 | 2.66 | 0.02 | 2.16 | 0.18 |
| 2.8 | 1.65 | 0.37 | 2.8 | - | - | 2.8 | 1.65 | 0.37 | 2.75 | 0.03 | 1.94 | 0.38 | 2.80 | 0.02 | 1.75 | 0.21 |
| 2.9 | - | - | 2.9 | - | - | 2.9 | - | - | 2.84 | 0.03 | 1.68 | 0.35 | - | - | - | - |
| 3.0 | 2.07 | 0.23 | 3.0 | - | - | 3.0 | 2.07 | 0.23 | 2.96 | 0.03 | 2.06 | 0.26 | 3.00 | 0.02 | 2.07 | 0.14 |
| 3.1 | - | - | 3.1 | - | - | 3.1 | - | - | 3.04 | 0.03 | 2.08 | 0.24 | - | - | - | - |
| 3.2 | 2.36 | 0.30 | 3.2 | - | - | 3.2 | 2.36 | 0.30 | 3.16 | 0.03 | 2.34 | 0.29 | 3.17 | 0.02 | 2.35 | 0.21 |
| 3.3 | - | - | 3.3 | - | - | 3.3 | - | - | - | - | - | - | - | - | - | - |
| 3.4 | - | - | 3.4 | - | - | 3.4 | - | - | - | - | - | - | - | - | - | - |
| 3.5 | 1.73 | 0.53 | 3.5 | - | - | 3.5 | 1.73 | 0.53 | 3.50 | 0.06 | 1.81 | 0.43 | 3.50 | 0.04 | 1.77 | 0.33 |

**Notes:** Each of the columns (1), (2), (C$_1$), (C$_2$), and (CC), shows the datasets plotted in the panels of Figures 5(a), 5(b), 5(c), 6, and 5(d), respectively.

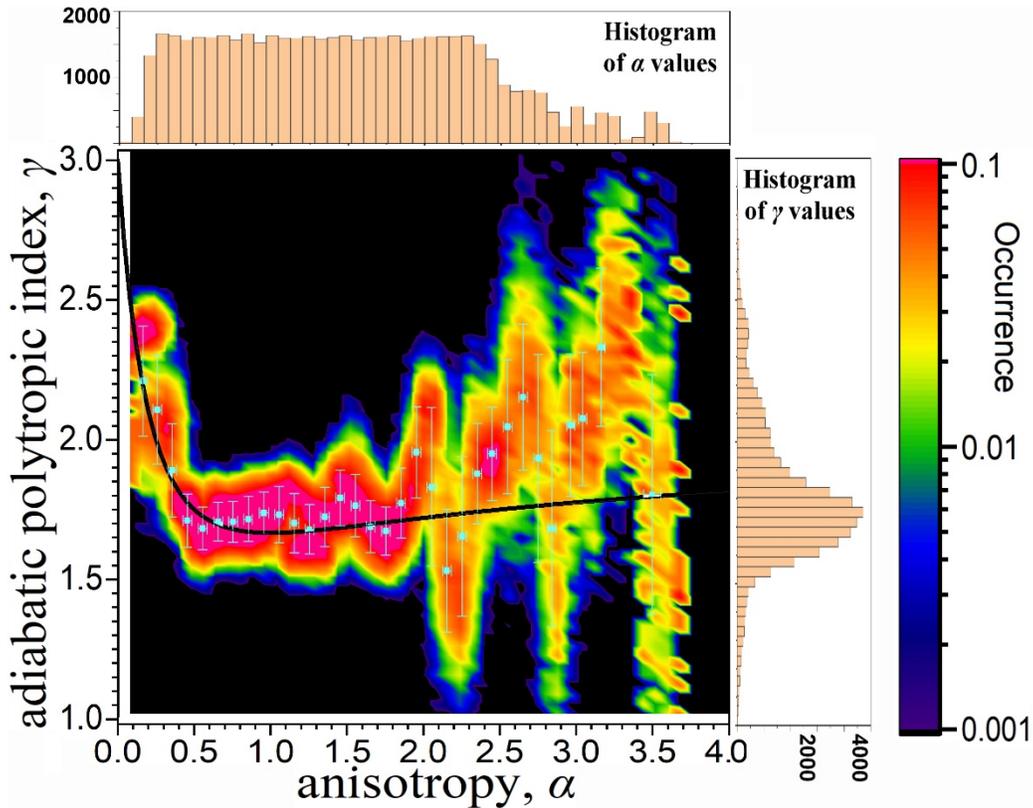

**Figure 6.** 2D-histogram of the enriched $N \times N_0$ dataset of $\gamma$ vs. $\alpha$ values. Also shown are the 1D histograms of $\alpha$ (upper) and $\gamma$ (right) panels. The 2D-histogram is normalized with the 1D histogram of $\alpha$ values. Also plotted are the homogenized dataset of polytropic indices $\{\gamma_i^{(C_2)} \pm \delta\gamma_i^{(C_2)}\}_{i=1}^{N_0}$ (light blue) and the modeled relationship in Eq.(18) (black curve).

*3.5. Comments*

The previous data and statistical analysis is performed to test the application of the theoretically developed model, in the solar wind proton plasma near 1 AU. The analysis showed that the derived theoretical relationship is in good agreement with observations. It also showed that the lowest and classical value of the adiabatic polytropic index, $\gamma=5/3$, occurs in the isotropic case, while anisotropic plasmas are characterized by higher polytropic indices.

Nicolaou et al. (2020) examined the large-scale variations of the proton plasma density and temperature within the inner heliosphere explored by Parker Solar Probe, and found polytropic indices higher than the adiabatic value, $\gamma > 5/3$, with a most frequent value of $\gamma \sim 2.7$. In particular, these authors discuss possible combinations of the effective kinetic degrees of freedom and the energy transfer terms, in order to interpret the short time scale fluctuations of the solar wind protons in the inner heliosphere, which seem to follow a polytropic law with $\gamma > 5/3$. Therefore, while the solar wind plasma near earth is mostly near adiabatic and thus isotropic (e.g., Livadiotis & Desai 2016), in smaller heliospheric distances the solar wind plasma is super-adiabatic, and thus, far from being isotropic.

## 4. Discussion: What's next?

### 4.1. Polytropic index vs. anisotropy, for plasma flows under a dynamical potential

In the previous sections (§2.2 and §3), we connected the adiabatic polytropic index with temperature anisotropy. Here we extend the theoretical analysis to derive the relationship between the anisotropy and any polytropic index − not just the adiabatic one; this is achieved by employing the already known connection between kappa and polytropic indices, for particles driven by potential energy (Livadiotis 2019b).

The connection of the polytropic index, adiabatic or non-adiabatic, is achieved through the effective dimensionality. In the case of the adiabatic polytropic index, we used the kinetic effective dimensionality, here noted by $d_u$, developed via the concept of correlation coefficient (as shown in §2.2 and Eq.(17)). In the case of a non-adiabatic polytropic index, the connection is with the positional effective dimensionality, which is noted by $d_r$. It was shown that the polytropic behavior of the plasma flow, driven by the dynamics of particle potential energy, is one-to-one equivalent with the description of kappa distributions of particle velocities/energies (e.g., Livadiotis 2019b); the connection between polytropic and kappa indices includes the positional dimensionality, $d_r$. Namely, we set

$$d_{\text{eff}} \to \begin{cases} d_u \to \gamma = 1 + 2\,d_u^{-1} \\ d_r \to \kappa + (\gamma-1)^{-1} = \tfrac{1}{2} + 2\,d_r \,, \end{cases} \tag{20}$$

where the effective dimensionality $d_{\text{eff}}$ is derived from the correlation coefficient in Eq.(16),

$$d_{\text{eff}} = \frac{(2\alpha+1)^2}{2\alpha^2+1} \,, \tag{21}$$

and interprets either the kinetic degrees of freedom, $d_{eff} = d_u$, leading to the adiabatic polytropic index according to Eqs.(17,18), or the positional degrees of freedom, $d_{eff} = d_r$, leading to the non-adiabatic polytropic index according to the relationship between kappa index and positional degrees of freedom (Livadiotis 2019b):

$$\kappa + (\gamma - 1)^{-1} = \tfrac{1}{2} + \tfrac{1}{b} d_r = \tfrac{1}{2} + 2 d_r \ . \tag{22}$$

Hence, we find

$$\kappa = \tfrac{1}{2} + \tfrac{1}{b} \frac{(2\alpha + 1)^2}{2\alpha^2 + 1} - (\gamma - 1)^{-1} \ , \tag{23}$$

thus, we have the non-adiabatic polytropic index expressed in terms of the anisotropy

$$\gamma = \frac{\dfrac{(2\alpha + 1)^2}{2\alpha^2 + 1} - b(\kappa - \tfrac{3}{2})}{\dfrac{(2\alpha + 1)^2}{2\alpha^2 + 1} - b(\kappa - \tfrac{1}{2})} \ , \tag{24a}$$

or vice-versa,

$$\alpha = \frac{1 \pm \sqrt{\tfrac{1}{2} b [\kappa - \tfrac{1}{2} + (\gamma - 1)^{-1}] \left\{ 3 - b[\kappa - \tfrac{1}{2} + (\gamma - 1)^{-1}] \right\}}}{b[\kappa - \tfrac{1}{2} + (\gamma - 1)^{-1}] - 2} \ . \tag{24b}$$

Figure 7 plots the polytropic index, $\nu$ and $\gamma$, as a function of the anisotropy $\alpha$, for two cases of attractive potentials, that is, a power-law with exponent $b=-1/2$ (e.g., interplanetary electric field potential, e.g., Cuperman & Harten 1971; Lacombe et al. 2002, Livadiotis 2018c; Nicolaou & Livadiotis 2019) (upper panels), and $b=2$ (e.g., centrifugal potential Meyer-Vernet et al. 1995; Livadiotis 2015a;c) (lower panels).

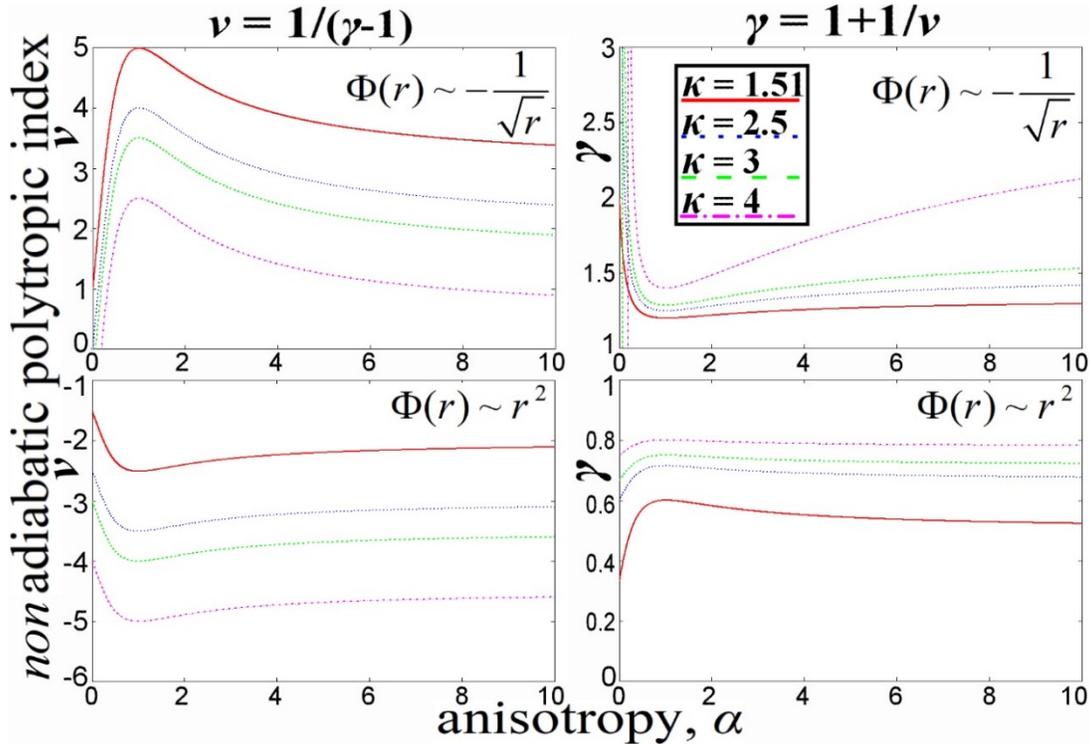

**Figure 7.** Relationship between non-adiabatic polytropic and kappa indices with anisotropy. The polytropic index, which is related with kappa distributions of particles with potential energy, Eq.(24a), is plotted as a function of the anisotropy, and for various kappa indices; the potential energy is a power-law with exponent $b=-1/2$ (e.g., interplanetary electric field potential) (upper panels), and $b=2$ (e.g., centrifugal potential) (lower panels).

*4.2. Correlation coefficients for anisotropic distributions with heterogeneous correlations*

There is a number of different types or models for the anisotropic kappa distributions that were frequently or rarely used in the past. In this paper, we have dealt with the most frequent one, expressed by Eq.(2a). The correlation coefficient, the effective dimensionality, and the adiabatic polytropic index were derived from considering only this certain type of kappa distributions. However, there are another two associated models that can be examined in a future analysis.

The model in Eq.(2a) characterizes correlated particles, for which the particle velocity components are homogeneously correlated, namely, the correlations between velocity components of each particle, are equal to the correlations between the same components of different particles. On the other hand, another − less frequently – model considers the case where the heterogeneous correlations are equal to zero; namely, there are no correlations between velocity components of each particle, and finite correlations between the same components of different particles. These two types of models were generalized by considering arbitrary homogeneous/heterogeneous correlations among the particles' velocity components. In particular, the generalized model mediates the mentioned two types of anisotropic kappa distributions, where the first considers equal correlations among particles velocity components, while the second considers zero correlation among different velocity components. (For more details, see: §4 in Livadiotis et al. 2020).

Below we present the three associated models of anisotropic kappa distributions. The correlation coefficient between two velocity components is symbolized by the related kappa index; we recall that the correlation coefficient is inversely proportional to the kappa index Eq.(14). The standard kappa index $\kappa$ corresponds to the correlation between the same velocity components of different particles, while the kappa index denoted by $\kappa^{\text{int}}$ corresponds to the correlation between the different velocity components of the each particle. Namely, we have:

- Homogeneous correlations for all velocity components, $\kappa^{\text{int}}=\kappa$,

$$P(u_{\parallel}, u_{\perp}; \theta_{\parallel}, \theta_{\perp}, \kappa) = [\pi (\kappa - \tfrac{3}{2})]^{-\frac{3}{2}} \cdot \frac{\Gamma(\kappa+1)}{\Gamma(\kappa - \tfrac{1}{2})} \cdot \theta_{\parallel}^{-1} \theta_{\perp}^{-2} \cdot \left[ 1 + \frac{1}{\kappa - \tfrac{3}{2}} \cdot \left( \frac{u_{\parallel}^2}{\theta_{\parallel}^2} + \frac{u_{\perp}^2}{\theta_{\perp}^2} \right) \right]^{-\kappa-1}, \tag{25a}$$

or, in terms of anisotropy $\alpha$,

$$P(\vec{u}; \theta, \alpha, \kappa; \kappa^{\text{int}} = \kappa) = \pi^{-\frac{3}{2}} \cdot (\kappa - \tfrac{3}{2})^{-\frac{3}{2}} \frac{\Gamma(\kappa+1)}{\Gamma(\kappa - \tfrac{1}{2})}$$
$$\times \alpha^{-1} [\tfrac{1}{3}(1+2\alpha)]^{\frac{3}{2}} \theta^{-3} \cdot \left[ 1 + \frac{1}{\kappa - \tfrac{3}{2}} \cdot \frac{1+2\alpha}{3\alpha\theta^2} \cdot (\alpha u_{\parallel}^2 + u_{\perp}^2) \right]^{-\kappa-1}. \tag{25b}$$

- Heterogeneous correlation (between parallel and perpendicular components) equal to zero, $\kappa^{\text{int}} \rightarrow \infty$,

$$P(u_{\parallel}, \vec{u}_{\perp}; \theta_{\parallel}, \theta_{\perp}, \kappa) = [\pi (\kappa - \tfrac{3}{2})]^{-\frac{3}{2}} \cdot \frac{\Gamma(\kappa)(\kappa - \tfrac{1}{2})}{\Gamma(\kappa - \tfrac{1}{2})} \cdot \theta_{\parallel}^{-1} \theta_{\perp}^{-2}$$
$$\times \left( 1 + \frac{1}{\kappa - \tfrac{3}{2}} \cdot \frac{u_{\parallel}^2}{\theta_{\parallel}^2} \right)^{-\kappa} \cdot \left( 1 + \frac{1}{\kappa - \tfrac{3}{2}} \cdot \frac{u_{\perp_x}^2 + u_{\perp_y}^2}{\theta_{\perp}^2} \right)^{-\kappa - \frac{1}{2}}, \tag{26a}$$

or

$$P(\vec{u};\theta,\alpha,\kappa;\kappa^{int}\rightarrow\infty)=[\pi(\kappa-\tfrac{3}{2})]^{-\frac{3}{2}}\cdot\frac{\Gamma(\kappa)(\kappa-\tfrac{1}{2})}{\Gamma(\kappa-\tfrac{1}{2})}\cdot\alpha^{-1}[\tfrac{1}{3}(1+2\alpha)]^{\frac{3}{2}}\theta^{-3}$$

$$\times\left(1+\frac{1}{\kappa-\tfrac{3}{2}}\cdot\frac{1+2\alpha}{3\alpha\theta^2}\cdot u_\perp{}^2\right)^{-\kappa-\frac{1}{2}}\cdot\left(1+\frac{1}{\kappa-\tfrac{3}{2}}\cdot\frac{1+2\alpha}{3\theta^2}\cdot u_\parallel{}^2\right)^{-\kappa}.$$

(26b)

- Heterogeneous correlations with arbitrary value, $\kappa^{int}<\infty$,

$$P(u_\parallel,u_\perp;\theta_\parallel,\theta_\perp,\kappa,\kappa^{int})=\mathrm{C}(\kappa^{int},\kappa)\cdot\theta_\perp{}^{-2}\theta_\parallel{}^{-1}$$

$$\times\left\{-1+\left(1+\frac{1}{\kappa-\tfrac{3}{2}}\cdot\frac{u_\perp{}^2}{\theta_\perp{}^2}\right)^{\frac{\kappa+\frac{1}{2}}{\kappa^{int}+\frac{1}{2}}}+\left(1+\frac{1}{\kappa-\tfrac{3}{2}}\cdot\frac{u_\parallel{}^2}{\theta_\parallel{}^2}\right)^{\frac{\kappa}{\kappa^{int}}}\right\}^{-\kappa^{int}-1}.$$

(27a)

or

$$P(\vec{u};\theta,\alpha,\kappa;\kappa^{int})=\mathrm{C}(\kappa^{int},\kappa)\cdot\alpha^{-1}[\tfrac{1}{3}(1+2\alpha)]^{\frac{3}{2}}\theta^{-3}$$

$$\times\left[-1+\left(1+\frac{1}{\kappa-\tfrac{3}{2}}\cdot\frac{1+2\alpha}{3\alpha\theta^2}\cdot u_\perp{}^2\right)^{\frac{\kappa+\frac{1}{2}}{\kappa^{int}+\frac{1}{2}}}+\left(1+\frac{1}{\kappa-\tfrac{3}{2}}\cdot\frac{1+2\alpha}{3\theta^2}\cdot u_\parallel{}^2\right)^{\frac{\kappa}{\kappa^{int}}}\right]^{-\kappa^{int}-1}.$$

(27b)

where the normalization constant is given by

$$\mathrm{C}(\kappa^{int},\kappa)^{-1}=\int_0^\infty\int_{-\infty}^{+\infty}\left[-1+\left(1+\frac{x_\perp{}^2}{\kappa-\tfrac{3}{2}}\right)^{\frac{\kappa+\frac{1}{2}}{\kappa^{int}+\frac{1}{2}}}+\left(1+\frac{x_\parallel{}^2}{\kappa-\tfrac{3}{2}}\right)^{\frac{\kappa}{\kappa^{int}}}\right]^{-\kappa^{int}-1}dx_\parallel\,2\pi x_\perp dx_\perp.$$

(27c)

Next, we deal with the case of anisotropic kappa distributions with zero heterogeneous correlations (between parallel and perpendicular components), $\kappa^{int}\rightarrow\infty$, given by Eq.(26b). Following steps similar to §2.2, we find that the correlation coefficient is again written as in Eq.(14), where the effective dimensionality is

$$d_{\mathrm{eff}}(\alpha)=\frac{1+4\alpha^2}{1+2\alpha^2},$$

(28)

In the general case, where we consider a $d$-D velocity space, the $d$-D velocity vector is decomposed to 1D parallel component and a ($d$-1)-D perpendicular component. Then, following again steps similar to §2.2, we find that the effective dimensionality is now given by

$$d_{\mathrm{eff}}(\alpha;d)=\frac{1+(d-1)^2\alpha^2}{1+(d-1)\alpha^2}.$$

(29a)

The effective dimensionality ranges between 1 and $d$-1 as expected. Namely, when $\alpha=0$, the perpendicular projection is suppressed and the distribution becomes 1-dimensional, appearing like a cigar; when $\alpha\rightarrow\infty$, the perpendicular projection is prevailed the parallel and the distribution becomes ($d$-1)-dimensional, appearing thus, like a ($d$-1)-D pie. However, the effective dimensionality of the distribution is never equal to the actual dimensionality of the $d$-D velocity space, even in the isotropic case, $\alpha=1$. The reason is the existence of unequal heterogeneous correlations among different velocity components; that is, homogeneous correlation between the perpendicular components, corresponding to $\kappa^{\mathrm{int}}=\kappa$, and zero correlation between perpendicular and parallel components, $\kappa^{\mathrm{int}}\to\infty$.

The adiabatic polytropic index, corresponding to the effective dimensionality of unequal heterogeneous correlations, is derived by following the steps in Eqs.(17,18), leading to

$$\gamma(\alpha;d)=1+2\cdot\frac{1+(d-1)\alpha^2}{1+(d-1)^2\alpha^2} \ , \tag{29b}$$

where it ranges from $\gamma=1$ for $\alpha=0$ to $\gamma=(d+1)/(d-1)$ for $\alpha\to\infty$, while it is $\gamma=(d^2+2)/[1+(d-1)^2]$ for $\alpha=1$. In the 3D isotropic case ($\alpha=1$) we find $\gamma=11/5=2.2$, that is, not a minimum, but slightly above the value of $\gamma=2$ for $\alpha\to\infty$. Figure 8 plots the effective dimensionality and the adiabatic polytropic index with respect to the anisotropy for both the distributions given in Eq.(25b) and Eq.(26b).

The data and statistical analysis performed in §3, showed that the theoretically developed model is in good agreement with the observations in the solar wind proton plasma. However, as shown in Figure 6, a possible deviation of the observations from the modeled relationship appears for large anisotropies; indeed, for $\alpha > 2$ the observed adiabatic polytropic index appears larger than its modeled value. This can be explained by the existence of heterogeneous correlations, expressed either as $\kappa^{\mathrm{int}}\to\infty$, or at least, as $\kappa < \kappa^{\mathrm{int}} < \infty$. It is possible, then, high anisotropies ($\alpha > 2$) to lead to deviations of the standard anisotropic kappa distributions from the formula in Eq.(26a) for $\kappa^{\mathrm{int}}=\kappa$ to the formula Eq.(26b) for $\kappa^{\mathrm{int}}\to\infty$, or to the formula Eq.(26c) for any finite $\kappa^{\mathrm{int}}$, i.e., $\kappa < \kappa^{\mathrm{int}}<\infty$.

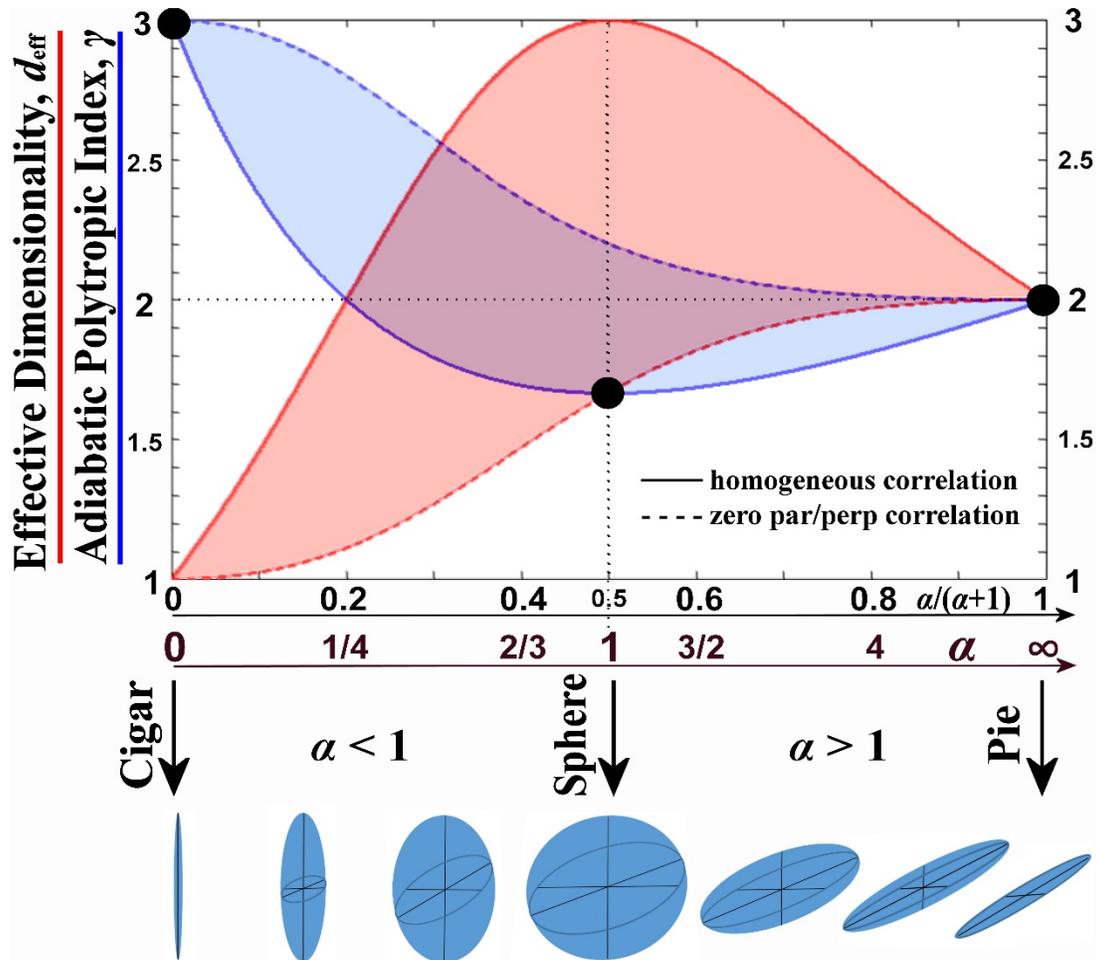

**Figure 8.** Similar to Figure 2, plotting the effective dimensionality $d_{eff}$ (red) and the respective adiabatic polytropic index $\gamma$ (blue) as a function of the anisotropy $\alpha$, derived from the cases of homogeneous distribution ($\kappa^{int}=\kappa$) shown in Eq.(25b) (solid line) and of heterogeneous with zero correlation between perpendicular and parallel components ($\kappa^{int}\rightarrow\infty$) shown in Eq.(26b) (dash line). All the in-between cases (transparent red/blue) correspond to other non-trivial correlations (different finite values of $\kappa^{int}$).

## 5. Conclusions

The paper showed a relationship between the polytropic index $\gamma$ and the temperature anisotropy $\alpha$ that may characterize space plasmas. In particular, the paper (i) developed a theoretical model that connects the adiabatic polytropic index and the temperature anisotropy; a similar connection will be developed for the non-adiabatic polytropic indices; and (ii) performed a data and statistical analysis in order to examine the validity of the theoretically developed model, in the case of the solar wind proton plasma near 1 AU; this was accomplished using datasets taken from Wind S/C.

The derived theoretical relationship found to be in good agreement with observations. According to the results of the analysis, the lowest and classical value of the adiabatic polytropic index, $\gamma=5/3$, occurs in the case of isotropic distributions, while higher indices characterize anisotropic plasmas. A possible deviation of the observations from the modeled relationship appears for large anisotropies (Figure 6); indeed, for $\alpha >2$ the observed average adiabatic polytropic index is larger than its modeled value. This high-anisotropy deviation can be explained by the existence of heterogeneous correlations. The basic model of anisotropic kappa distributions characterizes correlated particles, for which the particle velocity components are homogeneously correlated, namely, the correlations between different velocity components of each particle, are equal to the correlations between the same components of different particles. On the other hand, other, less frequently observed models, consider heterogeneous correlations among particle velocity components. (For more details, see: Livadiotis et al. 2020).

Therefore, the results showed that the adiabatic polytropic index can be larger than the standard value of $\gamma = 5/3$ in the case where the solar wind plasma is characterized by highly anisotropic distributions and/or when heterogeneous correlations exist among particle velocity components.

Nicolaou et al. (2020) examined the large-scale variations of the proton plasma density and temperature within the inner heliosphere explored by Parker Solar Probe, and found polytropic indices higher than the adiabatic value, $\gamma > 5/3$, with most frequent value $\gamma \sim 2.7$. In particular, these authors discuss possible combinations of the effective kinetic degrees of freedom and the energy transfer terms, in order to interpret the short time scale fluctuations of the solar wind protons in the inner heliosphere, which seem to follow a polytropic law with $\gamma > 5/3$. Therefore, while the solar wind plasma near earth is mostly adiabatic (e.g., Livadiotis & Desai 2016), and thus isotropic (Figure 6), in smaller heliospheric distances the solar wind plasma is super-adiabatic, and thus, far from being isotropic.

In summary, the paper results are outlined as follows:

- Derive the theoretical relationship between the adiabatic polytropic index and the anisotropy.
- Perform data and statistical analysis, showing that the theoretically developed model is in good agreement with the observations in the solar wind proton plasma.
- Show that the lowest and classical value of the adiabatic polytropic index, $\gamma = 5/3$, occurs in the isotropic case, while anisotropic plasmas (mostly with $\alpha > 2$) are characterized by higher polytropic indices.
- Show that in the adiabatic polytropic index is even larger for highly anisotropic distributions when heterogeneous correlations exist among particle velocity components (mostly with $\kappa < \kappa^{int} < \infty$).
- Development possible extensions of the theory considering (i) non-adiabatic polytropic behavior, and (ii) more general distributions.

**ORCID iDs**

G. Livadiotis https://orcid.org/0000-0002-7655-6019
G. Nicolaou https://orcid.org/0000-0003-3623-4928